\documentclass[aps,prd,twocolumn,superscriptaddress,showpacs,amsmath,amssymb]{revtex4}
\usepackage{graphicx,epsf}
\usepackage[]{latexsym}
\newcommand{\be}{\begin{eqnarray}}
\newcommand{\ee}{\end{eqnarray}}
\newcommand{\bra}[1]{\mbox{$\langle\, #1 \mid$}}

\newcommand{\ket}[1]{\mbox{$\mid #1\,\rangle$}}

\newcommand{\pro}[2]{\mbox{$\langle\, #1 \mid #2\,\rangle$}}

\renewcommand{\d}{\mbox{{\rm d}}}
\begin{document}
\title{Moving mirrors and black hole evaporation in non-commutative space-times}
\author{R. Casadio}
\email{casadio@bo.infn.it}
\affiliation{Dipartimento di Fisica, Universit\`a di Bologna,
I.N.F.N., Sezione di Bologna, via Irnerio 46, 40126 Bologna, Italy}
\author{P. H. Cox}
\email{phcox@tamuk.edu}
\affiliation{Department of Physics/Geosciences, Texas A\&M
University-Kingsville, M.S.C. 175, 700 University Blvd.,
Kingsville, TX 78363-8202}
\author{B. Harms}
\email{bharms@bama.ua.edu}
\affiliation{Department of Physics and Astronomy, The University
of Alabama, Box 870324, Tuscaloosa, AL 35487-0324, USA}
\author{O. Micu}
\email{micu001@bama.ua.edu}
\affiliation{Department of Physics and Astronomy, The University
of Alabama, Box 870324, Tuscaloosa, AL 35487-0324, USA}
\begin{abstract}
We study the evaporation of black holes in non-commutative
space-times. We do this by calculating the correction to the
detector's response function for a moving mirror in terms of the
noncommutativity parameter $\Theta$ and then extracting the number
density as modified by this parameter. We find that allowing space
and time to be non-commutative increases the decay rate of a black
hole.
\end{abstract}
\maketitle
\large
\section{Introduction}
\label{intro}
Non-commutativity is an intrinsic feature of quantum theories.
It is manifested in quantum mechanics in the phase-space
commutation relations
\be
\left[p_i,x_j\right] = i\,\hbar\,\delta_{ij}
\ ,
\ee
and in quantum field theory in the commutation relations of
creation and annihilation operators.
The idea that space-time coordinates do not commute arises
in string theory~\cite{witten,seiberg} and in the present search for
quantum gravity~\cite{moffat}, while Yang-Mills theories on
non-commutative spaces~\cite{witten1} appear in string theory
and M-theory.
\par
Non-commutative geometry is an old proposal~\cite{snyder} based on the
concept that there might exist a fundamental length in
the fabric of space-time.
For a parameter to be considered a fundamental length, it should respect
Lorentz invariance.
In order to preserve Lorentz invariance, one needs to include the
time coordinate among the non-commutative variables, but this is not
a trivial change. In fact, theories in which the time coordinate
is non-commutative seem to be acausal.
An example of such a theory is given in Ref.~\cite{seiberg},
where the authors study the effects of space-time non-commutativity
on the scattering of wave packets in the context of field theory.
In the same paper they show that in the formalism of string theory,
stringy effects cancel the acausal effects that appear in field theories.
Further, even if in the relativistic regime unitarity is violated, this
does not preclude the existence of a unitary low energy regime.
\par
Other motivations for non-commutativity come from D-brane
scenarios~\cite{witten,seiberg}.
Space non-commutativity appears when D-branes occur
in hyper-magnetic fields.
Time non-commutativity is generated similarly by nonzero
hyper-electric fields.
\par
Non-commutative space-time is defined in terms of space-time
coordinates $x^\mu$ ($\mu=1,2,\ldots,D$) which satisfy the following
commutation relations:
\be
\left[x^\mu,x^\nu\right]=-i\,\Theta^{\mu\nu}
\ .
\ee
$\Theta$ must be an antisymmetric Lorentz tensor~\cite{smailagic}.
As such, it can be transformed into a block-diagonal form:
\be
\hat\Theta^{\mu\nu}={\rm diag}
\left(\hat\Theta_1,\hat\Theta_2,\ldots,\hat\Theta_{D/2}\right)
\ ,
\ee
where
\be
\hat\Theta_i=\Theta_i\,
\left(
{\begin{array}{cc}
0 & 1
\\
[2ex] -1 &  0
\\
\end{array}}
\right)
\ .
\label{tens}
\ee
Since a coordinate reversal changes the sign of a $\Theta$, we can
without loss of generality require that all $\Theta_i$ be positive.
\par
In order to have full non-commutativity, one needs to work in a
space-time that has an even number of dimensions.
Then the $D=2\,d$ coordinates can be represented by $d$
two-vectors:
\be
\hat{x}^\mu&=&\left(\hat{x}^1,\hat{x}^2,\ldots,
\hat{x}^{2d-1},\hat{x}^{2d}\right)
\nonumber
\\
&=&
\left(\vec{\hat{x}}_1,\vec{\hat{x}}_2,\ldots,
\vec{\hat{x}}_{d}\right)
\ ,
\ee
with
$\vec{\hat{x}}_{i} \equiv \left(\hat{y}_{1i},\hat{y}_{2i}\right)$
being two-vectors in the $i$-th non-commutative plane that satisfy
\be
\left[\hat{y}_{1i},\hat{y}_{2i}\right]=i\,\Theta_i
\ .
\ee
\par
In a coherent state approach a set of commuting ladder operators
is constructed from non-commutative space-time coordinates
only~\cite{spallucci}.
We define the ladder operators for the $i$-th plane in the following
way,
\be
\!\!\!\!\! \hat a_i
&\!=\!\!& \frac{1}{\sqrt{2}}\left(\hat y_{1i}+i\,\hat y_{2i}\right)
 , \,
\hat a_i^{\dagger} =\frac{1}{\sqrt{2}}\left(\hat y_{1i}-i\,\hat y_{2i}\right)
.
\ee
These operators will then satisfy the canonical commutation rules
\be
\left[\hat a_i,\hat a_j^{\dagger}\right]=\delta_{ij}\,\Theta_i
\ .
\ee
Normalized ($\pro{\alpha}{ \alpha}=1$) coherent states can now be defined
for these operators as
\be
\ket{\alpha}=\prod_{i}\,\exp\left[
\frac{1}{\Theta_i}\,\left(
\overline{\alpha}_i\,\hat a_i-\alpha_i\,\hat a_i^{\dagger}\right)\right]
\ket{0}
\ ,
\ee
where $\ket{0}$ is a vacuum state, annihilated by all $\hat a_i$.
\par
Commutative coordinates are associated with the non-commuting ones
as their mean values over coherent states.
In this way, a non-commutative plane wave can be calculated using
Hausdorff decomposition, resulting in the following form
\begin{widetext}
\be
\bra{\alpha}
\exp\left[i\,\sum_{i=1}^{d}\,\left(
\vec{p}\cdot\vec{\hat{x}}\right)_i\right]
\ket{\alpha}=
\exp\left\{-\sum_{i=1}^{d}\left[
\frac{1}{4}\,\Theta_i (p_{1i}^{\;2}
+p_{2i}^{\;2})+i\left(\vec{p}\cdot\vec{{x}}\right)_i
\right]
\right\}
\ ,
\ee
\end{widetext}
where $p_{1i}$ and $p_{2i}$ are the momenta canonically conjugate
to the space-time coordinates.
Eq.~(10) then shows that a plane wave in the non-commutative
case will have damping factors proportional to $\Theta_i$ and that
we recover the usual form in the limit $\Theta_i\to 0$.
\par
Using this non-commutative form in the plane wave expansion of a
scalar field, we next study the creation of particles by a single
reflecting boundary (a moving mirror).
The mode solutions for a receding mirror are the same as the
late-time asymptotic modes for a ball of matter undergoing
gravitational collapse to form a black hole.
The Bogoliubov transformations for the two systems are almost
identical~\cite{B&D}.
The close connection between the mathematical descriptions of
these two systems implies that we can use the number density
distribution obtained from the detector response function for the
moving mirror to calculate the rate of decay of a black
hole~\cite{hawking}.
\section{moving mirror in Two-dimensional space-time}
We start with a brief introduction to the moving mirror
for the commutative case in two dimensions as treated in~\cite{B&D}.
We denote two-dimensional coordinates by $\vec x$
[that is, $\vec x=(t,x)$ or $(u,v)$ defined below],
so that the mirror is represented by a point that will move along
a trajectory
\be
\label{traj}
x &=& z(t) \ ,
\;
\left\{
\begin{array}{l}
|\dot{z}(t)|<1
\\
\\
z(t)=0 \ ,\quad t<0
\end{array} \  \right .
\ee
\par
In terms of light-cone coordinates $u=t-x$ and $v=t+x$,
a massless scalar field $\phi$, which satisfies the field equation
\be
\Box\phi=\frac{\partial^2\phi}{\partial{u}\,\partial{v}}=0
\ee
and the reflection boundary condition
\be
\phi\left(t,z(t)\right)=0
\ ,
\label{reflCond}
\ee
has the set of mode solutions
\be
u_k^{\mathrm in}=\frac{i}{\sqrt{4\,\pi\,\omega}}\,
\left[
e^{-i\,\omega\,{v}}
-e^{i\,\omega\,\left(2\,\tau_u-u\right)}
\right]
\ ,
\label{modesC}
\ee
where $\omega=|k|$ and $\tau_u$, the coordinate of the reflection point
of the ray, for the trajectory~(\ref{traj}), is the solution of
\be
\tau_u-z\left(\tau_u\right)=u
\ .
\ee
The scalar field (operator) $\hat\phi$ to the right of the mirror
can then be written in terms of the modes~(\ref{modesC})
as the usual superposition
\be
\hat \phi=\sum_{k>0}\left[\hat a_{k}\,u_{k}^{\rm in}
+\hat a_k^{\dagger}\,(u_{k}^{\rm in})^{*}\right]
\ .
\ee
\par
For $t<0$, we assume the mirror is at rest ($\tau_u=0$)
and the scalar field is in the vaccum state $\ket{0,{\rm in}}$.
The Wightman function,
\be
\!\!\!\!\!\!
D^{+}(\vec x;\vec x')&\!\!\equiv\!\!&
\bra{{\rm in},0}\hat \phi(\vec x)\,
\hat\phi(\vec x')\ket{0,{\rm in}}
\nonumber
\\
&\!\!=\!\!&
\sum_k\,u_k^{\rm in}(u,v)\,\left[u_k^{{\rm in}}(u',v')\right]^*
\ ,
\label{wightDef}
\ee
in this region becomes~\cite{B&D}
\be
D^{+}_<=
\frac{-1}{4\,\pi}\,
\ln
\left[
\frac{(u-u'-i\epsilon)(v-v'-i\epsilon)}
{(v-u'-i\epsilon)(u-v'-i\epsilon)}
\right]
,
\label{wightC<}
\ee
which exhibits the usual $i\epsilon$ prescription.
\par
As the mirror moves for $t>0$, it will mimic
a time-dependent background geometry,
such as that produced by a gravitational source.
If we define $p(u)=2\,\tau_u-u$, the Wightman
function in the vacuum and for a general trajectory
can be written as
\be
D^{+}_>\!=\!
\frac{-1}{4\pi}
\ln\!\left[\frac{(p(u)-p(u')-i\epsilon)(v-v'-i\epsilon)}
{(v-p(u')-i\epsilon)(p(u)-v'-i\epsilon)}\right]
.
\nonumber
\\
\label{wightC>}
\ee
From this non-trivial result, one in general expects a non-zero
detector response function for late times (formally, for
detector proper time $\tau\to\infty$),
\be
F(\omega)&\!\!\!=\!\!\!&\!
\int\!\!\!\!\!
\int_{-\infty}^{\infty}\!\!\!\!\!
\d\tau\,
\d\tau'\,
e^{-i\,\omega\,(\tau-\tau')}
D^{+}(\tau,x(\tau);\tau',x(\tau')\!)
\nonumber
\\
&\!\!\!\simeq\!\!\!&\! \int\!\!\!\!\!
\int_{-\infty}^{\infty}\!\!\!\!\! \d\tau\, \d\tau'\,
e^{-i\,\omega\,(\tau-\tau')}\,
D^{+}_>(\tau,x(\tau);\tau',x(\tau')\!) , \nonumber
\\
\label{F} \ee where $x=x(\tau)$ is the trajectory of the detector, which
we assume is adiabatically turned on, and later off, in the asymptotic future.
\par
A case of special interest is a mirror trajectory
with the asymptotic form
\be
\label{asymtraj}
z(t)\sim -t-Ae^{-2\kappa t}+B \ ,
\quad {\rm for} \  t\rightarrow\infty
\ ,
\ee
with $A$, $B$, and $\kappa$ positive constants.
For this type of mirror trajectory, null rays with $v<B$ reflect
from the mirror, rays with $v>B$ pass undisturbed, and the
ray $v=B$ is somewhat like a horizon.
A similar situation occurs when a star collapses
to a black hole.
The function $D^{+}_>$ can be written as a sum of
four terms.
If we choose the trajectory (\ref{asymtraj}) and the detector is
switched off at early times, the only non-zero contribution
will be from the term involving $p(u)-p(u')$, and
\be
p(u)\sim B-A\,e^{-\kappa\,(u+B)}
\quad
{\rm for}\ u\to\infty
\ .
\ee
For a detector that moves at constant velocity,
\be
x=x_0+w\,t \ , \label{dtraj}
\ee
the response function per unit of proper time $\tau$
is then given by
\be
\frac{\d F(\omega)}{\d\tau}
=\frac{1}{\omega\,\left(
e^{\omega/k_{\rm B}\,T}-1\right)}
\ ,
\label{dFC}
\ee
where the temperature $T$ is defined as
\be
k_{\rm B}\,T =
\frac{\kappa}{2\,\pi}\,\sqrt{\frac{1-w}{1+w}}
\equiv
\frac{\kappa\,\alpha}{2\,\pi}
\ ,
\label{TC}
\ee
and $k_{\rm B}$ is the Boltzmann constant.  We also introduce
here the Doppler-shift factor $\alpha$.
\par
It is important to note that the final result~(\ref{dFC})
does not depend on the details of the mirror and detector
trajectories, represented by the parameters $A$, $B$ and $x_0$.
The detector's velocity just enters as a kinematical (Doppler shift)
factor and the interesting physics is entirely contained in the
mirror's acceleration parameter $\kappa$.
We shall see that non-commutativity changes this scenario.
\section{Moving mirror in two-dimensional non-commutative space-time}
Since we are working in a two-dimensional space-time, there is only
one non-commutative plane and we will simply denote the
(positive) non-commutativity parameter by $\Theta$,
\be
\left[x^0,x^1\right]=-i\,\Theta
\ .
\ee
A set of modes analogous to those in Eq.~(\ref{modesC})
will have the form~\footnote{It is a subtle point how to treat
boundary conditions, such as that defining the interaction
between the scalar field and the mirror in
Eq.~(\ref{reflCond}), in non-commutative theories.
We simply assume here that the non-commutative length
$\sqrt{\Theta}$ is much shorter than the mirror's thickness
$\delta$ and their interaction is therefore not significantly
modified.
Mathematically, this means one should find the result for
$0<\Theta\ll\delta^2$ and then take the limit $\Theta\to 0$
{\em before\/} $\delta\to 0$.}
\be
u_{{\rm NC}}^{\rm in}=
\frac{i\,e^{-\frac{1}{2}\,\Theta\,\omega^2}}
{\sqrt{4\,\pi\,\omega}}\,
\left[e^{-i\,k\,v}-e^{-i\,k\,(2\,\tau_u-u)}\right]
\ ,
\label{modesNC}
\ee
again with $\omega=|k|$.
\par
As usual, we take the mirror to be at rest for negative times.
It starts moving at $t=0$, and the scalar field is in the state
$\ket{0,{\rm in}}$, void of particles, for negative times.
\subsection{The propagator}
The Wightman function for $t<0$ can be obtain by replacing the
commutative modes with the non-commutative ones~(\ref{modesNC})
in the sum~(\ref{wightDef}) and will therefore be given by
\begin{widetext}
\be
D^{+}_{NC<} & \!\!=\!\!&
\frac{1}{4\,\pi} \sum_{l=1}^{\infty}
\frac{1}{l}\,e^{-\frac{4\,\pi^2\,l^2}{L^2}\,\Theta}
\!
\left[e^{\frac{2\,i\,\pi\,l}{L}(v'-v)}
+e^{\frac{2\,i\,\pi\,l}{L}(u'-u)}
-e^{\frac{2\,i\,\pi\,l}{L}(v'-u)}
-e^{\frac{2\,i\,\pi\,l}{L}(u'-v)}
\right] \ ,
\ee
\end{widetext}
where we have used box normalization as a regulator, $L$ being the
size of the box (the continuum limit $L\to\infty$ will
be taken later in the computation) and $k=2\,\pi\,l/L$
is the discretized momentum.
\par
If we now expand the Gaussian term for small $\Theta$
as $\exp\left(-4\,\pi^2\,l^2\,\Theta/L^2\right)=
1-4\,\pi^2\,l^2\,\Theta/L^2
+\mathrm{O}(\Theta^2/L^4)$,
the non-commutative Wightman function can be easily calculated
to first order in $\Theta/L^2$,
\begin{widetext}
\be
D^{+}_{\rm NC<} \simeq
D^{+}_<
-\frac{\Theta}{4\,\pi}
\left[
\frac{1}{(u-u'-i\epsilon)^2}
+\frac{1}{(v-v'-i\epsilon)^2}
-\frac{1}{(u-v'-i\epsilon)^2}
-\frac{1}{(v-u'-i\epsilon)^2}\right]
,
\label{wightNC<}
\ee
for $t<0$, in which the first term is the commutative propagator,
Eq.~(\ref{wightC<}), and the remaining terms are
the contribution due to non-commutativity.
Analogously, for $t>0$, we obtain
\be
D^{+}_{\rm NC>}
\simeq
D^{+}_{>}
-\frac{\Theta}{4\,\pi}
\left[
\frac{1}{(p(u)-p(u')-i\epsilon)^2}
+\frac{1}{(v-v'-i\epsilon)^2}
-\frac{1}{(p(u)-v'-i\epsilon)^2}
-\frac{1}{(v-p(u')-i\epsilon)^2}\right]
.
\nonumber
\\
\label{wightNC>}
\ee
\end{widetext}
Note that the regulator $L$ does not appear in the
non-commutative part and the limit $L \to \infty$ can thus be
taken without difficulty.
\subsection{Detector response}
We want to evaluate the effect of non-commutativity on the detector
response function for the case that we discussed above.
\par
For negative times, it can be easily verified by substituting
Eq.~(\ref{wightNC<}) into Eq.~(\ref{F}) that an inertial detector which
is switched off for $t\to-\infty$, and which is moving along a trajectory
of the form~(\ref{dtraj}), will not record any particles.
\par
For $t>0$, we again consider the class of detector trajectories in
Eq.~(\ref{dtraj}), and estimate the contribution to the Wightman function
from non-commutativity,
\begin{widetext}
\be
\Delta F(\omega)\equiv
F_{\rm NC}(\omega)-F(\omega)
&\!\!\simeq\!\!&
-\frac{\Theta}{4\,\pi}
\int\!\!\!\!\!
\int_{-\infty}^{\infty}
\!\!\!
\d\tau\,
\d\tau'\,
e^{-i\,\omega\,(\tau-\tau')}
\left[\frac{1}{(p(u)-p(u')-i\epsilon)^2}
\right.
\nonumber
\\
&&
\phantom{-\frac{\Theta}{4\,\pi}}
\left.
+\frac{1}{(v'-v-i\epsilon)^2}
-\frac{1}{(v'-p(u)-i\epsilon)^2}
+\frac{1}{(v-p(u')-i\epsilon)^2}
\right]
\,  ,
\ee
\end{widetext}
in which $F(\omega)$ is the commutative part from Eq.~(\ref{F}).
\par
Three of the above terms from the non-commutative part will again
give zero contributions, and one finds
\begin{widetext}
\be
\Delta F(\omega)
\simeq
-\frac{\Theta}{4\,\pi}
\int\!\!\!\!\!
\int_{-\infty}^{\infty}
\frac{e^{-i\,\omega\,(\tau-\tau')}\,
\d\tau\,\d\tau'}{(p(u)-p(u')-i\epsilon)^2}
=
-\frac{\Theta\,e^{2\,\kappa\,B}}
{4\,\pi\,A^2\,e^{\kappa\,x_0}}
\int\!\!\!\!\!
\int_{-\infty}^{\infty}
\frac{e^{-i\,\omega\,(\tau-\tau')}\,\d\tau\,\d\tau'}
{\left(e^{-\kappa\,\alpha\,\tau'}
-e^{-\kappa\,\alpha\,\tau}-i\epsilon\right)^2
}
\,  .
\ee
\end{widetext}
where $\alpha$ is the Doppler shift factor, Eq.~(\ref{TC}).
It is now convenient to introduce the variables
$\Delta\tau=\tau-\tau'$ and $\xi=(\tau+\tau')/2$,
so that the non-commutative contribution to the detector's
response per unit time $\xi$ can be written as
\be
\!\!\!\!\!\!\!\!
\frac{\d{\Delta F}(\omega)}{\d\xi}
&\!\!\simeq\!\!&
-\frac{e^{2\,\kappa\,(B+\alpha\,\xi)}\,\Theta}
{16\,\pi\,A^2\,e^{\kappa\,x_0}}
\nonumber
\\
&&
\ \ \ \
\times
\int_{-\infty}^{\infty}
\frac{\d(\Delta\tau)\,e^{-i\,\omega\,\Delta\tau}}
{\sinh^2\left(\frac{1}{2}\,\kappa\,\alpha\,\Delta\tau-i\epsilon\right)}
\nonumber
\\
&\!\!=\!\!&
-\frac{e^{2\,\kappa\,(B+\alpha\,\xi)}\,\Theta}
{4\,\pi\,A^2\,e^{\kappa\,x_0}}
\nonumber
\\
&&
\times \!
\sum_{k=-\infty}^{\infty}
\int_{-\infty}^{\infty} \!
\frac{\d(\Delta\tau)\,e^{-i\,\omega\,\Delta\tau}}
{\left(\Delta\tau+i\,\frac{2\,\pi}{\kappa\,\alpha}\,k-i\epsilon\right)^2}
,
\ee
in which we use the identity
\be
\csc^2(\pi\,x)=
\frac{1}{\pi^2}\,
\sum_{n=-\infty}^{\infty}\left(x-n\right)^{-2}
\ee
to single out the poles in the complex $\Delta\tau$ plane.
By performing the contour integration and summing over $k$,
one finally obtains
\be
\frac{\d\Delta F(\omega)}{\d\xi}
\simeq
\frac{e^{2\,\kappa\,(B+\alpha\,\xi)}\,\Theta}
{2\,A^2\,e^{\kappa\,x_0}\,\kappa^2\,\alpha^2}\,
\frac{\omega}{e^{\frac{2\,\pi\,\omega}{\kappa\,\alpha}}-1}
\ .
\ee
\par
We can now add the contribution for the commutative
part and obtain the response per unit time
to first order in $\Theta$,
\be
\frac{\d F_{\rm NC}(\omega)}{\d\xi}
&\!\!\simeq\!\!&
\left(1+
\frac{e^{4\,\pi\,k_{\rm B}\,T\,(B/\alpha+\xi)}\,\omega^2\,\Theta}
{8\,\pi^2\,A^2\,e^{2\,\pi\,k_{\rm B}\,T\,x_0/\alpha}\,k_{\rm B}^2\,T^2}
\right)
\nonumber
\\
&&
\times
\frac{1}{\omega\,\left(e^{\omega/k_{\rm B}\,T}-1\right)}
\ ,
\label{dFNC}
\ee
where we have again used Eq.~({\ref{TC}).
Note that the result now depends on the details of the mirror
and detector trajectoies.
\section{Black hole decay rate}
The number of detected particles per mode and per unit time is
related to the detector's response function per unit time in
Eq.~(\ref{dFNC}) by \be N(\omega)=\omega\,\frac{\d F_{\rm
NC}(\omega)}{\d t} \ . \ee If we employ the analogy with the
Hawking effect, the temperature is related to the black hole
mass $M$ by the well known expression~\cite{hawking}
\be k_{\rm B}\,T=
\left(8\,\pi\,M\right)^{-1}
\ ,
\ee
(with $G=c=1$), and the rate at which the mass $M$ decreases is then
given by
\begin{widetext}
\be
\frac{\d M}{\d t}
\simeq
-\frac{1}{2\,\pi}\int_0^\infty
\left(1+
\frac{8\,M^2\,\omega^2\,e^{(B/\alpha+t)/2\,M}\,\Theta}
{A^2\,e^{x_0/4\,\alpha\,M}}
\right)\,
\frac{\Gamma(\omega)\,\omega\,\d\omega}
{\left(e^{8\,\pi\,M\,\omega}-1\right)}
\ ,
\ee
\end{widetext}
where $\Gamma$ is the grey-body factor and the integration is
over all available frequencies (formally, up
to infinity~\footnote{Of course,
energy conservation actually requires that modes with energy $\omega>M$
are not emitted (for the microcanonical approach to this issue, see
Refs.~\cite{micro}).}.)
\par
For the purpose of comparing with standard results, we will just
consider the simplest situation in which the decay takes place
through emission of spin zero massless particles for
which~\cite{page}
\be
\Gamma=\frac{{\mathcal A}}{\pi}\,\omega^2
\ ,
\ee
where ${\mathcal A}=16\,\pi\,M^2$ is the area of the black hole
horizon.
After we perform the integration over the whole range of energies,
the decay rate of the black hole will be
\be
\frac{\d M}{\d t}\simeq
-\frac{1}{7680\,\pi\,M^2}\,
\left(1
+\tilde\Theta\,e^{\frac{t}{2\,M}}\right)
\ ,
\ee
where we absorb in $\tilde\Theta$ the dependence on the
parameters $A$, $B$ and
$x_0$ which determine the mirror's trajectory.
\par
\begin{figure}[t]
\centering
\epsfxsize=2.8in
\epsfbox{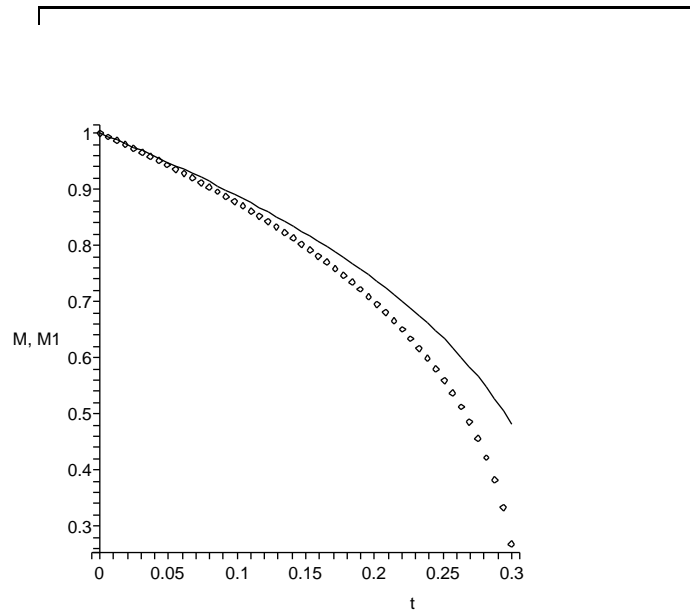}
%
\caption{Time evolution of the black hole mass in the commutative case
(solid line) compared with the non-commutative case (dotted line).
(Normalizations are arbitrary.)}
\label{Edist}
\end{figure}
Note that, contrary to the commutative case, the behavior now
depends significantly on these parameters.
For example, $x_0$ should be the detector's position for the simple
case $w=0$ (it appears that one does not learn anything more for
$w\not=0$) and one would naturally set $x_0\gg 2\,M$,
since the detector in the black hole case should stay sufficiently
away from the horizon.
This will of course damp the non-commutative correction exponentially.
Analogously, one may guess that $B\sim 2\,M$ ($\sim 1/T$), since that
is the asymptotic position of the mirror (the horizon), and it would
then follow that $\kappa\,B$ is just a number which can be simply
eliminated by rescaling $\Theta$ (to the same value for all black hole
evolutions).
However, $A$ represents a transient in the mirror's trajectory
which could be different for different time evolutions, thus allowing
to distinguish between different black hole evolutions, and that
it enters the final result is a novelty (the standard result does
not appear to depend on any transients to leading order in
perturbative field theory on the black hole background).
\par
We now calculate the mass of the black hole as a function of time
by integrating numerically the previous equation.
For the purpose of displaying a result, we arbitrarily set
$\tilde\Theta=0.1$,\footnote{This value is certainly too large to be
physically meaningful but does not change the qualitative behavior.}
and plot the time dependence of the black hole mass in Fig.~\ref{Edist}.
The graph shows that the black hole life-time is decreased if
space-time is non-commutative.
Of course, this result bears the same limitations as the original
Hawking's formula, that is, it is unreliable for small black hole
mass for which microcanonical corrections become
relevant~\cite{micro}.
\section{Discussion}
The form for the mathematical expression for the decay rate of
a black hole seems to depend strongly upon the assumptions made.
In most cases when a deviation from the approach used by Hawking
is made, e.g.~assuming the existence of large extra dimensions,
microcanonical versus the canonical
ensemble~\cite{micro}, or using the Randall-Sundrum brane-world
scenario~\cite{CG}, the decay rate of a black hole is reduced
compared to that of the Hawking result.
\par
The same conclusion seems to follow from regular black hole
solutions~\cite{regBH}, recently (re)discovered in the context
of gravity with a minimal length~\cite{nico} which is naturally
related to non-commutativity in space-time.
The increased decay rate for the case of non-commutative
space-time which we found therefore appears to be somewhat unusual.
It is possible that the analogy between the black hole and the
``kinematical'' model of the moving mirror cannot be simply carried
on to the non-commutative case, or that what we found ``adds'' to
the sort of effects obtained in the different approaches mentioned
above.
\par
Barring the above argument, our result may have interesting cosmological
implications for phenomena involving primordial black holes.
We do not
have a numerical estimate of the decrease in life-time of an evaporating
black hole, but the effect will be small since the parameter of
non-commutativity $\Theta$ is small in some sense.
Nevertheless, primordial black holes evaporating in non-commutative
space-time would have to be created at an early stage in the evolution
of the universe with an even larger mass than the $10^{15}$ grams
required in the commutative case in order to have survived to the
present epoch.
\end{document}